# Popular and/or Prestigious? Measures of Scholarly Esteem

**Ying Ding, Blaise Cronin**

{dingying, bcronin}@indiana.edu

School of Library and Information Science, Indiana University, Bloomington, IN 47405, USA

**Abstract**

Citation analysis does not generally take the quality of citations into account: all citations are weighted equally irrespective of source. However, a scholar may be highly cited but not highly regarded: popularity and prestige are not identical measures of esteem. In this study we define popularity as the number of times an author is cited and prestige as the number of times an author is cited by *highly* cited papers. Information Retrieval (IR) is the test field. We compare the 40 leading researchers in terms of their popularity and prestige over time. Some authors are ranked high on prestige but not on popularity, while others are ranked high on popularity but not on prestige. We also relate measures of popularity and prestige to date of Ph.D. award, number of key publications, organizational affiliation, receipt of prizes/honors, and gender.

## 1. Introduction

In the arts, as in other spheres of creative and sporting endeavor, popularity should not be confused with prestige. Topping the bestseller lists will not greatly affect an author's chances of winning the Nobel Prize for literature, nor is a Hollywood blockbuster that breaks box office records likely to land the *Palme d'Or* at Cannes. Similarly, impressive auction house sale prices are no guarantee that MoMA or Tate Modern will acquire an artist's work. Popular appeal and peer esteem are not synonymous, as sociologists of culture and others have noted (e.g., English, 2005). Things, of course, are not that different in the symbolic capital markets of academia (Bourdieu, 1988; Cronin, 1999; Cronin & Shaw, 2002).

Bollen, Rodriguez and Van de Sompel (2006) distinguished between scholarly popularity and prestige. They compared *journal* rankings resulting from a weighted PageRank metric (prestige) with those obtained using the impact factor (popularity) (see also Franceschet, 2009). In this paper we focus primarily on *authors* rather than journals. The popularity of a social actor (artist, pianist, scholar) can be defined as the total number of endorsements (acclaim, applause, citation) received from all other actors and prestige as the number of endorsements coming specifically from experts (see Bollen, Rodriguez & Van de Sompel, 2006, p. 2). Bibliometrically, popularity can be operationalized as the number of times an author is cited (endorsed) in total, and prestige as the number of times an author is cited by *highly* cited papers. A scholar may be popular but popularity does not necessarily equate with prestige, though on occasion there may well be a strong positive correlation between the two measures. For a thoroughgoing review of the concepts of prestige, prestige hierarchies and prestige scales, as well as related notions such as esteem, charisma, hierarchy and status, the reader is referred to Wegener (1992).

In the vernacular, it is not how often one is cited but by whom; that is to say, a citation from a Fellow of the Royal Society would for most of us carry more weight than one from a doctoral student. Likewise, a citation coming from an obscure paper probably would not be granted the same weight as a citation from a groundbreaking article (Bollen, Rodriguez & Van de Sompel, 2006; Maslov & Redner, 2008). Here we take the *quality* of citing articles into consideration in assessing the standing of researchers, using information retrieval as our test site.

In the present study, the popularity of a researcher is measured by the number of times he is cited by all papers in the same dataset; the prestige of a researcher by the number of times he is cited by *highly* cited papers in that dataset. Popularity and prestige are differentiated on the basis of the presumptive quality of citations. We show how scholars' popularity and prestige rankings change over time. We also explore the relationship between popularity and prestige and variables such as date of Ph.D. degree award, receipt of honors/prizes, number of key publications, organizational affiliation, and gender. The paper is organized as follows. Section 2 discusses related work on citation analysis and research evaluation. Section 3 describes the methods we used to calculate popularity and prestige. Section 4 analyzes changes in scholars' popularity and prestige rankings over time. Section 5 links popularity and prestige with other variables. In Section 6 we summarize our findings and suggest possible future work.

## 2. Related Work

Quantitative measures of research impact have been used since the early 20th century (Garfield, 1999). Cason and Lubotsky (1936) employed journal-to-journal citation analysis to measure the dependence of journals on each other. Pinski and Narin (1976) developed a citation-based technique to measure the influence of scientific journals, subfields, and fields. They calculated the eigenvalue of a journal cross-citing matrix as a size-independent influence weight for journals. Impact factors have been used to determine the standing of journals (Garfield, 1999; Bordons, Fernandez & Gomez, 2002; Harter & Nisonger, 1997; Nederhof, Luwel & Moed, 2001), and the same principle has been used to measure the impact of web pages (Smith, 1999; Thelwall, 2001). The h-index and variants thereon have been employed to assess the performance of researchers (Hirsch, 2005; Cronin & Meho, 2006; Sorenson, 2009). Other more or less novel approaches to citation analysis continue to emerge (e.g., Redner, 1998; Jin, Liang, Rousseau & Egghe, 2007; Sidiropoulos, Katsaros & Manolopoulos, 2007).

Straightforward counting—the number of times a particular author, paper, journal, institution, country has been cited—is the most basic approach. Riikonen and Vihinen (2008) stress the importance of simple citation counting having examined the effects of assigning differential weights to citations. There are also more advanced techniques to determine a scholar's influence on a particular field or intellectual community, for example, author co-citation analysis (e.g., White & McCain, 1998), social network analysis (Newman, 2001; Yan & Ding, 2009), and PageRank (Ding, Yan, Frazho & Caverlee, 2009).

Recently, for instance, Sorensen (2009) applied citation analysis to post-1984 research on Alzheimer's Disease. Based on data extracted from PubMed and Thomson Reuters' Web of Science, the top 100 Alzheimer's investigators were identified and their h-indexes calculated. Sorensen then highlighted those scientists on his list who had won either or both of the two most prestigious Alzheimer's research awards. Riikonen and Vihinen (2008) examined the productivity and impact of more than 700 biomedical researchers in Finland from 1966 to 2000. Their study showed that actual publication and citation counts were better indicators of the scientific contribution of researchers, disciplines, or nations than impact factors. Cronin and Meho (2007) explored the relationship between researchers' creativity (production of key papers) and professional age in the field of information science, but they, like others, did not take into account the quality of citing articles in their analysis.

Pinski and Narin (1976) proposed giving greater weight to citations coming from a prestigious journal than to citations from a peripheral one, an approach also suggested by Kochen (1974). Habibzadeh and Yadollahie (2008) granted greater weight to citations if the citing journal had a higher impact factor than that of the cited journal and then calculated the weighted impact factor to better measure the quality of journals. Bollen, Rodriguez, and Van de Sompel (2006) proposed a weighted PageRank algorithm to obtain a metric of prestige for journals, and found significant discrepancies between PageRank and impact factor. They defined popular journals as those cited frequently by journals with little prestige, and prestigious journals as those with citations coming from highly influential journals. Popular journals normally have a high impact factor but a low weighted PageRank, while prestigious journals have a low impact factor but a high weighted PageRank. Bollen et al. argue that the impact factor is a measure of popularity not of prestige and in so doing they have challenged the status quo (Al-Awqati, 2007). It is also worth noting that although researchers have begun to take account of the differential coverage of databases used in large-scale citation analysis (e.g., Meho & Yang, 2007), they continue to ignore the variable quality of citing articles. In an effort to address this deficiency we here use weighted citation counts as a means of distinguishing between scholarly popularity and prestige.

The basic units of measurement in bibliometrics are authors, papers, and journals. Straightforward citation analysis is a very convenient but also somewhat crude method: the strengths and limitations of the Journal Impact Factor, for instance, have been debated extensively and reviewed thoroughly by Bensman (2007). Most studies do not distinguish between scholarly popularity (reflected in raw citation counts) and prestige (reflected in weighted citation counts). The difference between prestige and popularity at the journal level has been little addressed in the literature; notable exceptions are an early paper by Pinski and Narin (1976) and more recently a detailed proposal by Bollen, Rodriguez and Van de Sompel (2006). Very few researchers have applied these kinds of approach to the author and paper levels. Here, we describe in detail how weighted citation counting at the author level can be applied in order to differentiate between scholarly prestige and popularity.

# 3. Methods

<u>Data collection</u>

We chose information retrieval as our test field as both of us have some familiarity with the domain and the actors. This is an interdisciplinary field, one that brings together scholars from information science and computer science in particular. It is also a field that draws upon techniques and tools from a number of other areas. Our sample contains many individuals who are recognizably mainstream researchers in IR (e.g., Harman, Robertson, Saracevic) and others who are associated with more or less cognate fields (e.g., Chen, Kohonen, Stonebraker).

Papers and their cited references were harvested from Web of Science (WoS) for the period 1956 to 2008. Search strategies were based on the following terms (including plurals and variants) which were determined by checking Library of Congress Subject Headings and consulting several domain experts: INFORMATION RETRIEVAL, INFORMATION STORAGE and RETRIEVAL, QUERY PROCESSING, DOCUMENT RETRIEVAL, DATA RETRIEVAL, IMAGE RETRIEVAL, TEXT RETRIEVAL, CONTENT BASED RETRIEVAL, CONTENT-BASED RETRIEVAL, DATABASE QUERY, DATABASE QUERIES, QUERY LANGUAGE, QUERY LANGUAGES, and RELEVANCE FEEDBACK. In total, 15,370 papers (henceforth the IR paper dataset) with 341,871 cited references (henceforth the IR cited references dataset) were collected. The citation records comprised first author, year, source, volume, and page number. The dataset is split into four time periods: phase 1 (1956-1980), phase 2 (1981-1990), phase 3 (1991-2000), and phase 4 (2001-2008).

Measures of Popularity and Prestige

We measured the popularity of a researcher by the number of citations he received over time. For example, if researcher A was cited 50 times by papers published prior to 1980, his popularity for that period was 50. We measured a researcher's prestige by the number of citations he received from highly cited papers. For example, if researcher A received 5 citations from highly cited papers published prior to 1980, his prestige score for that period was 5 (see Figure 1).

Popularity of a researcher = Number of times cited by all papers

Prestige of a researcher = Number of times cited by highly cited papers

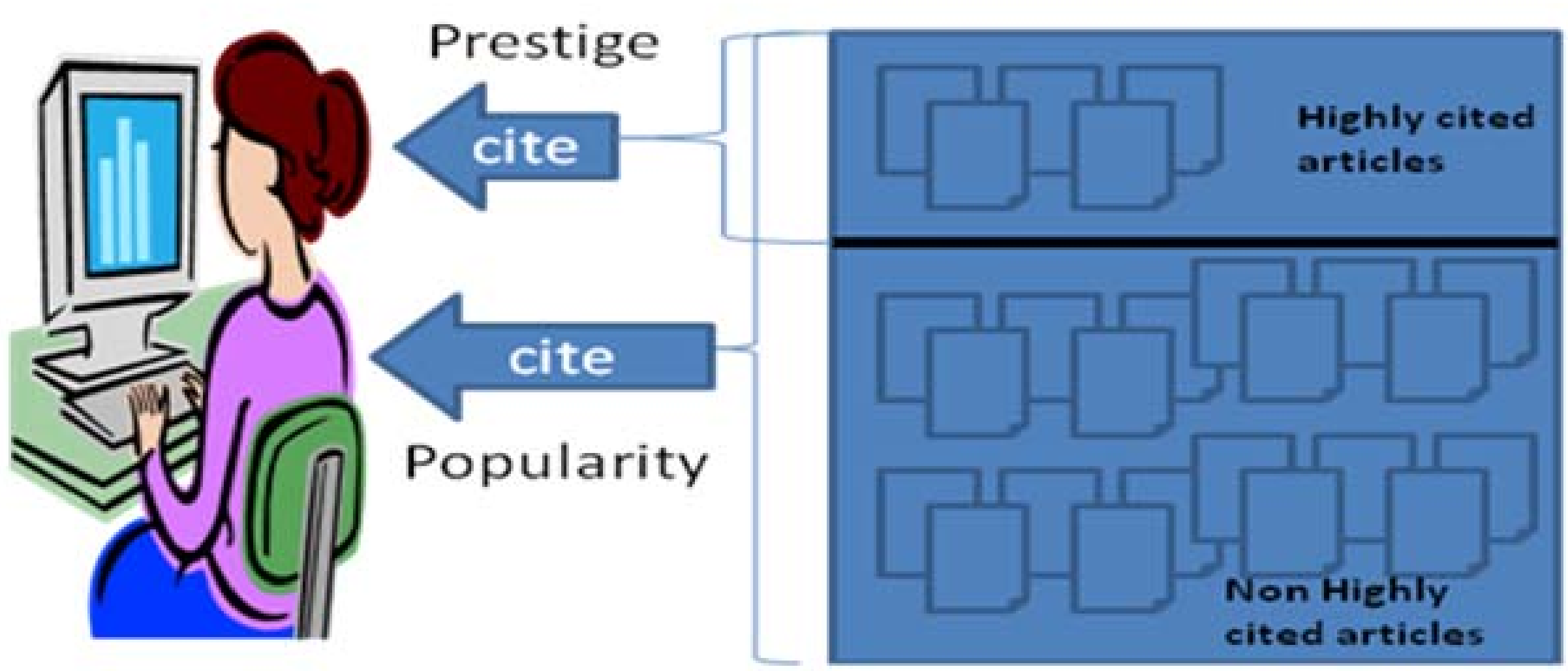

Figure 1. Popularity and prestige

Prestige calculation

*Step 1: Identify highly cited papers from the IR cited references dataset*

We identified a subset of highly cited papers from the IR cited references dataset for each time period. The subset contains roughly 20% of the total citations for each period: 2,379 highly cited papers (papers cited more than once) for 1956-1980, 4,243 (papers cited more than twice) for 1981-1990, 24,487 (papers cited four or more times) for 1991-2000, and 46,209 (papers cited five or more times) for 2001-2008. We sought to maintain the same ratio (roughly 20%) for each period for the sake of comparability, given that the time periods contain very different numbers of citations. For example, if we had defined highly cited papers as papers cited four or more times, we would have ended up with only 75 records for 1956-1980 but 23,487 for 1991-2000. Moreover, papers cited four or more times in 1956-1980 may be qualitatively different than those cited equivalently in 1991-2000; as the number of publications grows exponentially, the probability of citation increases.

*Step 2: Match highly cited papers against the IR paper dataset*

The first author name, publication year, volume and beginning page fields were used to match the highly cited papers against the IR paper dataset. Ultimately, 85 matches were recorded for 1956-1980, 136 for 1981-1990, 478 for 1991-2000, and 875 for 2001-2008.

Step 3: *Collect cited references in the matched papers and store them in the core cited references datasets*

We collected 1,603 cited references from the 85 highly cited papers for 1956-1980; 3,388 from the 136 papers for 1981-1990; 18,928 from the 478 papers for 1991-2000, and 35,305 from the 875 papers for 2001-2008.

Step 4: *Calculate the number of times each author has been cited in the core cited references datasets*

The prestige rankings of authors for the period 2001-2008 were calculated based on 35,305 cited references in the core cited references dataset—the number of times authors have been cited by the highly cited papers in this period. The process was idential for the three earlier time periods. Figure 2 illustrates the steps involved in generating measures of prestige.

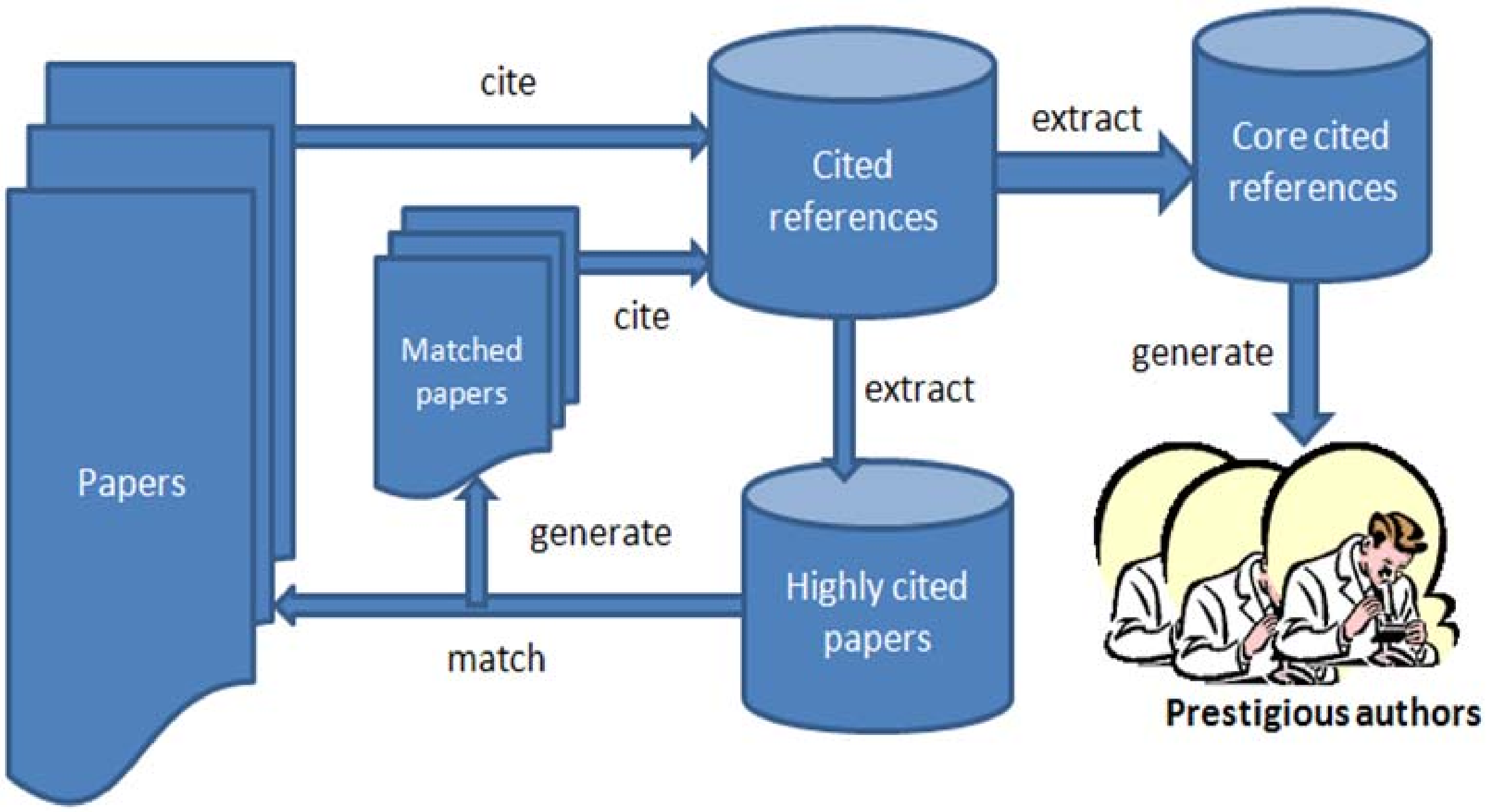

Figure 2. Calculating measures of prestige

## 4. Results and discussion

Dynamics of popularity

The left side of Table 1 shows the top 40 ranked IR authors in terms of popularity for each of the four time bands. Unsurprisingly, it is hard to maintain a continuous presence in the top 40 for fifty plus years . Many authors appeared once (e.g., Ingwersen P [NA-104-**19**-14193], Tahani V [232-**37**-939-2793]), several twice (e.g., Date CJ [389-**14-33**-389], Stonebraker M [1480-**15-14**-167], Borgman CL [NA-**18-12**-88]), and a few thrice (e.g., Cooper WS [**6-11-20**-121], Yu CT [**17-9-44**-431], Bates MJ [901-**39-9-38**]). Four were continuously present (marked in bold in Table 1): Salton G (1927-1995), Van Rijsbergen CJ, Robertson SE, and Jones KS (1935-2007). Each of these authors has made fundamental contributions to the field; at the risk of over-simplifying, the SMART system, theoretical models of IR, probabilistic searching model, and inverse document frequency respectively. Moreover, three won the Gerard Salton Award, named after the doyen of the field who, coincidentally, ranked top in both prestige and popularity across all four time periods.

Turnover is only to be expected. There are new entrants such as Spink A (NA-NA-**22-6**), Flickner M (NA-NA-**37-11**), Chen HC (NA-NA-**37-36**), Rui Y (NA-NA-194-**2**), Baeza-Yates R (NA-NA-593-**4**), and Smith JR (NA-NA-49-**5**). Flickner M, Rui Y, Baeza-Yates R and Smith JR currently work in industry (Yahoo!, IBM, Microsoft respectively), while Spink A and Chen HC are in academia. Most received their Ph.D. in the 1990s (three did not have a terminal degree), and typically spent 10 to 20 years working in the area before reaching the upper echelons (see Cronin & Meho [2007] on timelines of creativity in information science). Some authors' rankings are declining, for example, McCarn DB (**22**-175-2559-19223) and Doyle LB (**40**-196-577-4805). Some have left the field, retired, or died: Frome J (**9**-NA-NA-NA), Kent A (**4**-610-875-4822), Williams ME (**5**-46-555-2188) and Janda K (**32**-NA-NA-NA), for example.

Table 1. Top 40 ranked authors based on popularity and prestige

| | Popularity | | | | Prestige | | | |
|---|---|---|---|---|---|---|---|---|
| **R** | **1956-1980** | **1981-1990** | **1991-2000** | **2001-2008** | **1956-1980** | **1981-1990** | **1991-2000** | **2001-2008** |
| 1 | **SALTON G**<br>(1-1-1-1) | **SALTON G**<br>(1-1-1-1) | **SALTON G**<br>(1-1-1-1) | **SALTON G**<br>(1-1-1-1) | **SALTON G**<br>(1-1-1-1) | **SALTON G**<br>(1-1-1-1) | **SALTON G**<br>(1-1-1-1) | **SALTON G**<br>(1-1-1-1) |
| 2 | LANCASTER FW<br>(2-13-28-256) | **ROBERTSON SE**<br>(12-2-6-3) | ABITEBOUL S<br>(NA-52-2-7) | RUI Y<br>(NA-NA-194-2) | **JONES KS**<br>(2-3-6-10) | **ROBERTSON SE**<br>(10-2-3-4) | BELKIN NJ<br>(NA-12-2-2) | BELKIN NJ<br>(NA-12-2-2) |
| 3 | CLEVERDON CW<br>(3-37-81-256) | BOOKSTEIN A<br>(25-3-30-277) | BELKIN NJ<br>(206-10-3-8) | **ROBERTSON SE**<br>(12-2-6-3) | **LANCASTER FW**<br>(3-11-12-23) | **JONES KS**<br>(2-3-6-10) | **ROBERTSON SE**<br>(10-2-3-4) | SARACEVIC T<br>(71-64-7-3) |
| 4 | KENT A<br>(4-610-875-4822) | **VANRIJSBERGEN CJ**<br>(8-4-4-16) | **VANRIJSBERGEN CJ**<br>(8-4-4-16) | BAEZA-YATES R<br>(NA-NA-593-4) | **COOPER WS**<br>(4-10-10-16) | **VANRIJSBERGEN CJ**<br>(10-4-5-6) | CROFT WB<br>(NA-6-4-7) | **ROBERTSON SE**<br>(10-2-3-4) |
| 5 | WILLIAMS ME<br>(5-46-555-2188) | RADECKI T<br>(101-5-156-1664) | SARACEVIC T<br>(64-51-5-13) | SMITH JR<br>(NA-NA-49-5) | **BOOKSTEIN A**<br>(5-5-9-40) | **BOOKSTEIN A**<br>(5-5-9-40) | **VANRIJSBERGEN CJ**<br>(10-4-5-6) | SPINK A<br>(NA-NA-26-5) |
| 6 | COOPER WS<br>(6-11-20-121) | CROFT WB<br>(136-6-7-39) | **ROBERTSON SE**<br>(12-2-6-3) | SPINK A<br>(NA-NA-22-6) | MARON ME<br>(6-13-27-80)<br>**CLEVERDON CW**<br>(6-20-23-39)<br>SWETS JA<br>(6-20-104-376) | CROFT WB<br>(NA-6-4-7) | **JONES KS**<br>(2-3-6-10) | **VANRIJSBERGEN CJ**<br>(10-4-5-6) |
| 7 | **JONES KS**<br>(7-7-15-21) | **JONES KS**<br>(7-7-15-21) | CROFT WB<br>(136-6-7-39) | ABITEBOUL S<br>(NA-52-2-7) | | YU CT<br>(20-7-16-52) | SARACEVIC T<br>(71-64-7-3) | CROFT WB<br>(NA-6-4-7) |
| 8 | **VANRIJSBERGEN CJ**<br>(8-4-4-16) | CODD EF<br>(31-8-17-128) | ULLMAN JD<br>(NA-20-8-191) | BELKIN NJ<br>(206-10-3-8) | | RADECKI T<br>(NA-8-32-97) | BATES MJ<br>(424-38-8-11) | SMITH JR<br>(NA-NA-144-8) |
| 9 | FROME J<br>(9-NA-NA-NA) | YU CT<br>(17-9-44-431) | BATES MJ<br>(901-39-9-38) | VOORHEES EM<br>(NA-253-65-9) | CUADRA CA<br>(9-214-45-68) | ZADEH LA<br>(92-9-35-62) | **BOOKSTEIN A**<br>(5-5-9-40) | HARMAN D<br>(NA-NA-18-9) |
| 10 | CUADRA CA<br>(10-321-248-1493)<br>BOURNE CP<br>(10-1014-2285-6666) | BELKIN NJ<br>(206-10-3-8) | HARMAN D<br>(NA-421-10-19) | SMEULDERS AWM<br>(NA-NA-3524-10) | **ROBERTSON SE**<br>(10-2-3-4)<br>**VANRIJSBERGEN CJ**<br>(10-4-5-6) | **COOPER WS**<br>(4-10-10-16) | **COOPER WS**<br>(4-10-10-16) | **JONES KS**<br>(2-3-6-10) |
| 11 | | COOPER WS<br>(6-11-20-121) | CHANG SK<br>(1005-62-11-65) | FLICKNER M<br>(NA-NA-37-11) | | **LANCASTER FW**<br>(3-11-12-23)<br>BELKIN NJ<br>(NA-12-2-2) | BORGMAN CL<br>(NA-17-11-14) | BATES MJ<br>(424-38-8-11) |
| 12 | **ROBERTSON SE**<br>(12-2-6-3)<br>MIKHAILOV AI<br>(12-222-NA-NA) | ZADEH LA<br>(40-12-58-30) | BORGMAN CL<br>(NA-18-12-88) | JAIN AK<br>(NA-5595-64-12) | WILLIAMS ME<br>(12-47-272-952) | | **LANCASTER FW**<br>(3-11-12-23) | **SWANSON DR**<br>(19-15-13-12) |
| 13 | | LANCASTER FW<br>(2-13-28-256) | FALOUTSOS C<br>(NA-244-13-7615) | SARACEVIC T<br>(64-51-5-13) | DATTOLA RT<br>(13-42-604-1242)<br>DOYLE LB<br>(13-42-98-210)<br>REES AM<br>(13-827-53-75)<br>FUGMANN R<br>(13-NA-1161-1326)<br>BOURNE CP<br>(13-69-2567-792)<br>HAWKINS DT<br>(13-464-569-2153) | MARON ME<br>(6-13-27-80)<br>CODD EF<br>(31-13-21-94) | **SWANSON DR**<br>(19-15-13-12) | RUI Y<br>(NA-NA-196-13) |
| 14 | VICKERY BC<br>(14-97-367-1887) | DATE CJ<br>(389-14-33-389) | STONEBRAKER M<br>(1480-15-14-167) | SWAIN MJ<br>(NA-NA-72-14) | | | FOX EA<br>(NA-26-14-37) | PENTLAND A<br>(NA-NA-114-14)<br>BORGMAN CL<br>(NA-17-11-14) |
| 15 | MARON ME<br>(15-19-115-417) | STONEBRAKER M<br>(1480-15-14-167) | **JONES KS**<br>(7-7-15-21) | FALOUTSOS C<br>(NA-244-13-7615) | | **SWANSON DR**<br>(19-15-13-12) | BLAIR DC<br>(NA-84-15-35) | |
| 16 | HAWKINS DT<br>(16-69-824-2102) | BUELL DA<br>(NA-16-359-955) | KIM W<br>(NA-64-16-306) | **VANRIJSBERGEN CJ**<br>(8-4-4-16) | | HARPER DJ<br>(NA-16-74-141) | YU CT<br>(20-7-16-52) | **COOPER WS**<br>(4-10-10-16) |
| 17 | YU CT<br>(17-9-44-431) | GARFIELD E<br>(18-17-100-153) | CODD EF<br>(31-8-17-128) | FUHR N<br>(NA-584-32-17) | | BORGMAN CL<br>(NA-17-11-14)<br>**HARTER SP**<br>(36-17-22-21) | ABITEBOUL S<br>(NA-NA-17-35) | FUHR N<br>(NA-NA-30-17) |
| 18 | SUMMIT RK<br>(18-361-4183-NA)<br>GARFIELD E<br>(18-17-100-153) | BORGMAN CL<br>(NA-18-12-88) | BANCILHON F<br>(NA-68-18-925) | JANSEN BJ<br>(NA-NA-1044-18) | | | HARMAN D<br>(NA-NA-18-9) | FALOUTSOS C<br>(NA-NA-48-18) |
| 19 | | MARON ME<br>(15-19-115-417) | INGWERSEN P<br>(NA-104-19-14193) | HARMAN D<br>(NA-421-10-19) | **SWANSON DR**<br>(19-15-13-12)<br>LUHN HP<br>(19-38-153-302)<br>YU CT<br>(20-7-16-52) | CHAMBERLIN DD<br>(353-19-127-323) | FIDEL R<br>(NA-111-19-26) | INGWERSEN P<br>(NA-131-24-19) |
| 20 | CHERNYI AI<br>(20-135-NA-NA) | ULLMAN JD<br>(NA-20-8-191) | COOPER WS<br>(6-11-20-121) | MANJUNATH BS<br>(NA-NA-269-20) | | SWETS JA<br>(6-20-104-376)<br>HOLLAAR LA<br>(323-20-584-NA)<br>ODDY RN | ELLIS D<br>(NA-689-20-20) | ELLIS D<br>(NA-689-20-20) |
| 21 | LUHN HP<br>(21-1547-321-459) | BLAIR DC<br>(NA-21-21-129)<br>WILLETT P | BLAIR DC<br>(NA-21-21-129) | **JONES KS**<br>(7-7-15-21) | | | CODD EF<br>(31-13-21-94) | FLICKNER M<br>(NA-NA-383-21)<br>**HARTER SP** |

| | | | | | | | | | |
|---|---|---|---|---|---|---|---|---|---|
| 22 | TAUBE M<br>(22-509-1267-5875) | (NA-21-106-174) | SPINK A<br>(NA-NA-22-6)<br>ELLIS D<br>(NA-107-22-71) | PENTLAND A<br>(NA-NA-45-22) | | ROCCHIO JJ<br>(22-28-82-71)<br>HILLMAN DJ<br>(22-349-459-920)<br>BROOKES BC<br>(22-47-90-217)<br>VICKERY BC<br>(22-117-312-444)<br>MARCUS RS<br>(22-42-82-340)<br>STANDERA O<br>(22-NA-NA-NA)<br>SUMMIT RK<br>(22-410-1294-3612)<br>HEINE MH<br>(22-58-452-1440)<br>KATTER RV<br>(22-1064-423-1289) | (36-20-27-86)<br>**CLEVERDON CW**<br>(6-20-23-39) | **HARTER SP**<br>(36-17-22-21) | (36-17-22-21) |
| 23 | MCCARN DB<br>(22-175-2559-19223) | HARTER SP<br>(106-23-41-96)<br>KRAFT DH<br>(158-23-284-242) | | KOHONEN T<br>(NA-316-87-23) | | | | **CLEVERDON CW**<br>(6-20-23-39) | VOORHEES EM<br>(NA-790-103-23)<br>**LANCASTER FW**<br>(3-11-12-23)<br>SCHAMBER L<br>(NA-NA-46-23) |
| 24 | FUGMANN R<br>(24-1229-938-4620) | | FOX EA<br>(NA-30-24-233) | JOACHIMS T<br>(NA-NA-1860-24) | | | TAHANI V<br>(570-24-219-674)<br>MACLEOD IA<br>(70-24-206-1079) | INGWERSEN P<br>(NA-131-24-19) | |
| 25 | SWETS JA<br>(25-85-718-2540)<br>BOOKSTEIN A<br>(25-3-30-277) | MARCUS RS<br>(32-25-258-4790)<br>SWANSON DR<br>(27-25-26-94) | MARCHIONINI G<br>(NA-261-25-62) | BRIN S<br>(NA-NA-690-25)<br>VOORHEES E<br>(NA-1033-368-26) | | | | STONEBRAKER M<br>(NA-50-25-58) | |
| 26 | | | SWANSON D<br>(27-25-26-94) | | | | GARFIELD E<br>(NA-26-246-286)<br>FOX EA<br>(NA-26-14-37) | SPINK A<br>(NA-NA-26-5) | FIDEL R<br>(NA-111-19-26) |
| 27 | SWANSON DR<br>(27-25-26-94)<br>GOFFMAN W<br>(27-520-1715-1217)<br>MOOERS CN<br>(27-412-6146-7225) | MACLEOD IA<br>(87-27-315-3678) | HULL R<br>(NA-333-27-729) | AGRAWAL R<br>(NA-245-57-27) | | | | ODDY RN<br>(36-20-27-86)<br>MARON ME<br>(6-13-27-80) | MA WY<br>(NA-NA-204-27) |
| 28 | | SHNEIDERMAN B<br>(139-28-89-107)<br>MAIER D<br>(NA-28-29-418) | LANCASTER FW<br>(2-13-28-256) | WANG JZ<br>(NA-NA-2528-28) | | | ROCCHIO JJ<br>(22-28-82-71)<br>SMEATON AF<br>(NA-28-41-64) | | SWAIN MJ<br>(NA-NA-266-28) |
| 29 | | | MAIER D<br>(NA-28-29-418) | DEERWESTER S<br>(NA-3303-107-29) | | | | LOSEE RM<br>(NA-NA-29-111) | NIBLACK W<br>(NA-NA-108-29) |
| 30 | BORKO H<br>(30-294-NA-NA)<br>CODD EF<br>(31-8-17-128) | BERNSTEIN PA<br>(NA-30-120-529)<br>FOX EA<br>(NA-30-24-233)<br>MEADOW CT<br>(979-30-60-511)<br>HARPER DJ<br>(319-30-434-1162) | BOOKSTEIN A<br>(25-3-30-277) | ZADEH LA<br>(40-12-58-30) | | | DATE CJ<br>(513-30-42-203)<br>NOREAULT T<br>(912-30-122-510)<br>NEGOITA CV<br>(880-30-270-712)<br>IDE E<br>(31-30-129-167) | FUHR N<br>(NA-NA-30-17)<br>MARKEY K<br>(NA-243-30-88) | JAIN AK<br>(NA-NA-829-30) |
| 31 | | | GRAEFE G<br>(NA-367-31-99)<br>FUHR N<br>(NA-584-32-17) | MA WY<br>(NA-NA-169-31) | | MINKER J<br>(31-69-348-349)<br>CODD EF<br>(31-13-21-94)<br>BELLO F<br>(31-NA-NA-NA)<br>IDE E<br>(31-30-129-167)<br>PRYWES NS<br>(31-212-3886-9650) | | | CHANG SK<br>(NA-130-32-31) |
| 32 | REES AM<br>(32-450-353-1886)<br>MARCUS RS<br>(32-25-258-4790)<br>NEWMAN SM<br>(32-NA-NA-NA)<br>LANCASTER FW<br>(32-13-28-257)<br>OCONNOR J<br>(32-134-704-NA)<br>JANDA K<br>(32-NA-NA-NA)<br>HILLMAN DJ<br>(32-790-2308-17457) | | | CARSON C<br>(NA-NA-1071-32) | | | | CHANG SK<br>(NA-130-32-31)<br>RADECKI T<br>(NA-8-32-97) | CHEN HC<br>(NA-NA-35-32) |
| 33 | | | DATE CJ<br>(389-14-33-389) | BUCKLEY C<br>(NA-215-53-33) | | | | | MARCHIONINI G<br>(NA-NA-44-33) |
| 34 | | CHAMBERLIN DD<br>(302-34-681-5409)<br>ODDY RN<br>(119-34-141-5130)<br>DOSZKOCS TE<br>(93-34-199-2631) | FIDEL R<br>(NA-80-34-67) | PORTER MF<br>(NA-60-113-34) | | | KRAFT DH<br>(199-34-76-161)<br>PORTER MF<br>(NA-34-79-82)<br>WONG E<br>(95-34-157-346)<br>WILLETT P<br>(NA-34-72-99) | ULLMAN JD<br>(NA-216-34-137) | BUCKLEY C<br>(NA-394-55-34) |
| 35 | | | ELMASRI R<br>(NA-443-35-480)<br>BERTINO E<br>(NA-164-35-220) | YANG Y<br>(NA-4749-182-35) | | | | ZADEH LA<br>(92-9-35-62)<br>CHEN HC<br>(NA-NA-35-32)<br>CHEN H<br>(NA-NA-35-41)<br>WONG SKM<br>(NA-177-35-49) | BLAIR DC<br>(NA-84-15-35)<br>ABITEBOUL S<br>(NA-NA-17-35) |
| 36 | | | | CHEN HC<br>(NA-NA-37-36) | | KNUTH DE<br>(36-351-361-352)<br>MCCARN DB<br>(36-86-611-1684)<br>PADIN ME<br>(36-275-1359-NA)<br>**HARTER SP**<br>(36-17-22-21)<br>ODDY RN<br>(36-20-27-86)<br>THOMPSON DA<br>(36-799-1151-6685)<br>BORKO H<br>(36-935-556-667)<br>MARTIN TH<br>(36-69-395-1345)<br>RUBINOFF M<br>(36-NA-NA-NA)<br>AUGUSTSON JG<br>(36-208-1392-7791)<br>MEADOW CT<br>(36-47-42-55)<br>OCONNOR J<br>(36-236-150-669) | | | |
| 37 | | CLEVERDON CW<br>(3-37-81-256)<br>TAHANI V<br>(232-37-939-2793) | FLICKNER M<br>(NA-NA-37-11)<br>CHEN HC<br>(NA-NA-37-36) | HAWKING D<br>(NA-NA-294-37) | | | | | FOX EA<br>(NA-26-14-37) |
| 38 | | | | BATES MJ<br>(901-39-9-38) | | | LUHN HP<br>(19-38-153-302)<br>DOSZKOCS TE<br>(121-38-49-192)<br>BATES MJ<br>(424-38-8-11)<br>BUELL DA<br>(NA-38-96-234) | | KUHLTHAU CC<br>(NA-NA-56-38) |
| 39 | KESSLER MM<br>(39-534-1652-1052) | SMEATON AF<br>(NA-39-70-143)<br>BATES MJ<br>(901-39-9-38)<br>REISNER P<br>(NA-39-285-2193)<br>WONG E<br>(199-39-549-5157) | BEERI C<br>(NA-129-39-363) | LAWRENCE S<br>(NA-NA-397-39)<br>CROFT WB<br>(136-6-7-39) | | | | SNODGRASS R<br>(NA-71-39-184)<br>HULL R<br>(NA-NA-39-204) | **CLEVERDON CW**<br>(6-20-23-39) |
| 40 | MARTIN TH<br>(40-315-3464-NA)<br>BERNIER CL<br>(40-683-6305-NA)<br>DOYLE LB<br>(40-196-577-4805)<br>ZADEH LA<br>(40-12-58-30) | | CERI S<br>(NA-96-40-168)<br>HARTER SP<br>(106-23-41-96) | | | | | | **BOOKSTEIN A**<br>(5-5-9-40) |

Note: dd-dd-dd-dd: rank in phase 1- rank in phase 2 - rank in phase 3 - rank in phase 4. The authors marked in bold were continuously ranked in the top 40 for the entire period.

Information retrieval is a dynamic field. Only four authors were ranked in the top 40 for the entire period (Salton G, Jones KS, Van Rijsbergen CJ and Robertson SE). Among the top 40 ranked authors in phase 1, 16 kept their ranking in phase 2, 10 in phase 3, and 5 in phase 4. Among the top 40 in phase 2, 19 maintained their ranking in phase 3, and 8 in phase 4. In the case of phase 3, 14 of the top 40 kept their ranking in phase 4 (see the left side of Table 2 and Figure 3). Very roughly speaking, 40% of the authors in the top 40 were new entrants in each phase.

Table 2. Persistently popular and prestigious authors

| | Popularity | | | | Prestige | | | |
|---|---|---|---|---|---|---|---|---|
| | 1956-1980 | 1981-1990 | 1991-2000 | 2001-2008 | 1956-1980 | 1981-1990 | 1991-2000 | 2001-2008 |
| Phase 1 | 43 | 16 (37%) | 10 (23%) | 5 (12%) | 47 | 18 (38%) | 14 (30%) | 10 (21%) |
| Phase 2 | | 42 | 19 (45%) | 8 (19%) | | 41 | 21 (59%) | 15 (37%) |
| Phase 3 | | | 41 | 14 (34%) | | | 40 | 26 (65%) |

Note: The numbers in cells represent authors who maintained their ranking among the top 40 for successive phases. Ties in rank mean that *N* can exceed 40.

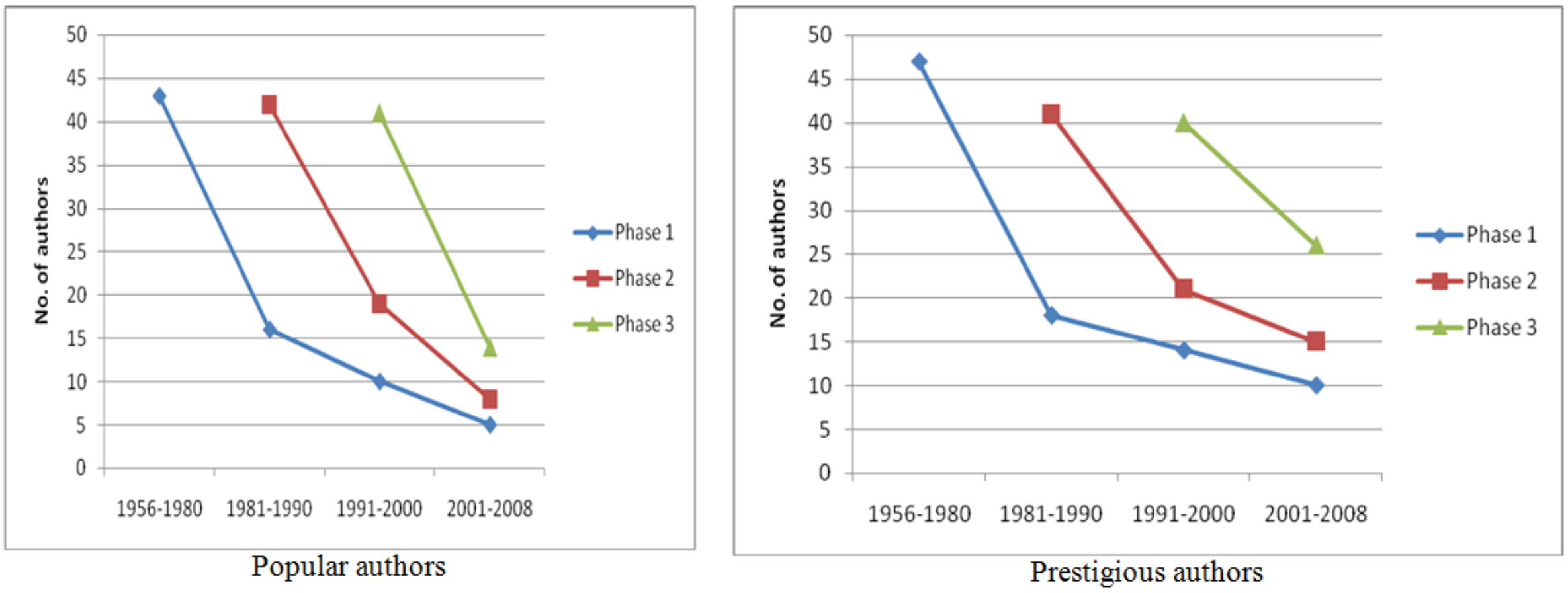

Figure 3. Persistently popular and prestigious authors

Shifting measures of esteem

The right side of Table 1 shows the top 40 authors ranked in terms of prestige. Ten had a continuous presence. This group included the four authors who were continuously ranked in the top 40 for popularity. The six other individuals (and their broadly defined areas of expertise) were: Lancaster FW for IR evaluation, Cooper WS for IR evaluation, Bookstein A for indexing theory, Swanson DR for medical IR, Cleverdon CW for IR evaluation, and Harter SP for probabilistic indexing. Some authors maintained their membership of the top 40 cohort for 10 years (e.g., Hillman DJ [**22**-349-459-920], Harper DJ [NA-**16**-74-141], Tahani V [570-**24**-219-674]), some for 20 years (e.g., Marcus RS [**22-42**-82-340], Luhn HP [**19-38**-153-303], Radecki T [NA-**8-32**-97], Ellis D [NA-689-**20-20**]), and a few for 30 years (e.g., Belkin NJ [NA-**12-2-2**], Fox EA [NA-**26-14-37**], and Codd EF [**31-13-21**-94]). Some stars were rising (e.g., Chen HC [NA-NA-**35-32**], Spink A [NA-NA-**26-5**], Harman D [NA-NA-**18-9**], Fuhr N [NA-NA-**30-17**]), while others were fading (e.g., Summit RK [**22**-410-1294-3612], Hawkins DT [**13**-464-569-2153], Padin ME [**36**-275-1359-NA]). Some names disappeared from the rankings (e.g., Bello F [**31**-NA-NA-NA], Rubinoff M [**36**-NA-NA-NA], Standera O [**22**-NA-NA-NA]).

Overall, the prestige rankings were more stable than the popularity rankings. Ten authors were continuously ranked within the top 40 for prestige (see the right side of Table 1, names in bold). Of the top 40 ranked authors in phase 1, 18 featured in phase 2, 14 in phase 3, and 10 in phase 4. Of the top 40 authors in phase 2, 21 maintained a presence in phase 3, and 15 in phase 4. Of the top 40 in phase 3, 26 maintained a presence in phase 4 (see the right sides of Table 2 and Figure 3). As a general rule, once an author is ranked high on prestige, i.e., is highly cited by important IR researchers, he tends to maintain his ranking for some time.

Popularity vs. Prestige

Popularity and prestige exist in the following possible relations:

- High popularity and high prestige

- High popularity and low prestige
- Low popularity and high prestige
- Low popularity and low prestige

Gerard Salton is a singularity in that he is consistently ranked highest in terms of both prestige and popularity. (The February 1996 issue of the *Journal of the American Society for Information Science* contains an In Memoriam that captures the nature of the man and his contributions.) Most of the top 10 ranked authors score highly in both the popularity and prestige stakes, such as Roberston SE (popularity rank: **12-2-6-3** vs. prestige rank: **10-2-3-4**), Jones SK (popularity rank: **7-7-15-21** vs. prestige rank: **2-3-6-10**), Van Rijsbergen CJ (popularity rank: **8-4-4-16** vs. prestige rank: **10-4-5-6**), while others have relatively low popularity and low prestige (within the top 40 ranked authors), such as Martin TH (popularity rank: **40**-315-3464-NA vs. prestige rank: **36**-69-395-1345). There are those whose rankings diverge. For example, people with high prestige rank but low popularity rank or the converse. For the period 2001-2008 there are many such cases: Croft WB (prestige rank **7**, popularity rank 39), Borgman CL (prestige rank **14**, popularity rank 88), Ingwersen P (prestige rank **19**, popularity rank 46), Marchionini G (prestige rank **33**, popularity rank 62); Maron ME (prestige rank 80, popularity rank 417); and Yu CT (prestige rank 52, popularity rank 431). These authors attract a relatively high number of citations from highly cited papers and a relatively low number of citations from non-highly cited papers. Conversely, some authors attract a relatively high number of citations from non-highly cited papers and a relatively small number of citations from highly cited papers. A large number of citations coming from non-highly cited papers will boost an author's popularity rank. There are several such cases for the years 2001-2008 (see Table 1).

Table 3. Popularity and prestige

| Ranks | 1956-1980 | 1981-1990 | 1991-2000 | 2001-2008 |
|---|---|---|---|---|
| No. 1-10 | 7 | 7 | 7 | 5 |
| No. 11-20 | 2 | 3 | 1 | 2 |
| No. 21-30 | 0 | 1 | 0 | 1 |
| No. 31-40 | 4 | 3 | 1 | 2 |

Note: Numbers in cells represent authors who were ranked in the top 40 for both popularity and prestige

Table 3 shows the number of authors ranked within the top 40 for both popularity and prestige across the four time periods. Many leading researchers were found among the top 10 in both categories across all four time periods. However, the popularity and prestige rankings of the researchers in ranks 11-40 differ appreciably. For example, the number of authors who were ranked high on both categories and across all time periods dropped from approximately 65% in ranks 1-10 to 20% in ranks 11-20, to 5% in ranks 21-30, and to 25% in ranks 31-40.

Validity

We tested the validity of the popularity and prestige ranks by comparing them with the rankings obtained by adding the impact factors of the journals in which the citing articles were published as weights to the raw citation counts. We limited our examination to 2001-2008, as this period contained the largest number of papers and citations. The Spearman correlation coefficient shows that prestige correlates weakly with popularity ($r=0.563$, $p<0.01$: See Table 5). Popularity, on the other hand, correlates strongly with impact factor ($r=0.939$, $p<0.01$: See Table 5), which confirms the findings of Bollen, Rodriguez and Van de Sompel (2006), namely, that the journal impact factor measures popularity rather than prestige. It can be inferred that prestige and popularity ranks measure slightly different dimensions of peer esteem. Figure 4 shows the scatter plots of these three different rankings, which underscores the point.

Table 4. Rankings for three measures of esteem (2001-2008)

| Authors | Prestige Rank | Popularity Rank | Impact Factor Rank |
|---|---|---|---|
| SALTON G | 1 | 1 | 1 |
| BELKIN NJ | 2 | 6 | 7 |
| SARACEVIC T | 3 | 13 | 12 |
| ROBERTSON SE | 4 | 3 | 3 |
| SPINK A | 5 | 6 | 4 |
| VANRIJSBERGEN CJ | 6 | 16 | 8 |
| CROFT WB | 7 | 39 | 23 |
| SMITH JR | 8 | 5 | 5 |
| HARMAN D | 9 | 19 | 14 |
| JONES KS | 10 | 21 | 18 |
| BATES MJ | 11 | 38 | 56 |
| SWANSON DR | 12 | 93 | 24 |
| RUI Y | 13 | 2 | 2 |
| PENTLAND A | 14 | 22 | 30 |
| BORGMAN CL | 14 | 88 | 106 |
| COOPER WS | 16 | 121 | 109 |
| FUHR N | 17 | 17 | 13 |
| FALOUTSOS C | 18 | 15 | 19 |
| INGWERSEN P | 19 | 46 | 59 |
| ELLIS D | 20 | 71 | 68 |
| FLICKNER M | 21 | 11 | 17 |
| HARTER SP | 21 | 95 | 98 |
| VOORHEES EM | 23 | 9 | 6 |
| LANCASTER FW | 23 | 256 | 336 |
| SCHAMBER L | 23 | 108 | 111 |
| FIDEL R | 26 | 67 | 70 |
| MA WY | 27 | 31 | 37 |
| SWAIN MJ | 28 | 13 | 20 |
| NIBLACK W | 29 | 60 | 91 |
| JAIN AK | 30 | 12 | 15 |
| CHANG SK | 31 | 65 | 99 |
| CHEN HC | 32 | 36 | 27 |
| MARCHIONINI G | 33 | 62 | 67 |
| BUCKLEY C | 34 | 33 | 31 |
| BLAIR DC | 35 | 128 | 119 |
| ABITEBOUL S | 35 | 7 | 26 |
| FOX EA | 37 | 233 | 281 |
| KUHLTHAU CC | 38 | 113 | 87 |
| CLEVERDON CW | 39 | 256 | 360 |
| BOOKSTEIN A | 40 | 276 | 207 |

Table 5. Correlations among various measures of esteem

| Spearman's rho | Prestige | Popularity | Impact Factor |
|---|---|---|---|
| Prestige | 1 | | |
| Popularity | 0.563 | 1 | |
| Impact Factor | 0.681 | 0.939 | 1 |

Note: Two-tailed Spearman correlation with significance at the 0.01 level

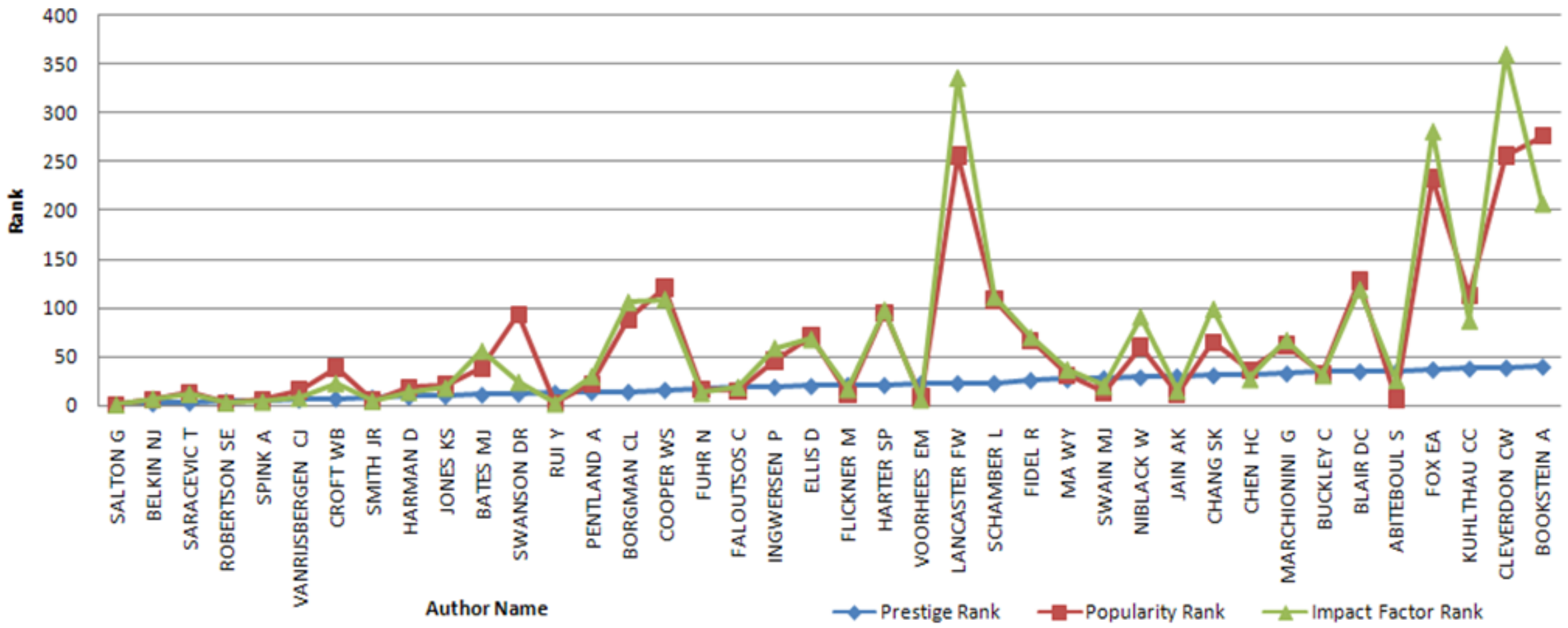

Figure 4. Scatter plots of the rankings based on prestige, popularity, and impact factor

## 5. Popularity, prestige, and other indicators of esteem

Table 6 shows the top 40 most highly cited/most popular authors from 1956 to 2008 along with related professional information: date of Ph.D. award, degree granting institution, institutional affiliation, major awards, service to the ACM SIGIR conferences, and an indication of authors' key contributions to the field. Almost all of the top 40 authors either work or have worked at leading universities (e.g., the University of California at Berkeley, University of Chicago, Stanford University) or research labs (e.g., IBM, Microsoft, Yahoo!). Twenty-five of these organizations are in the USA, 6 in the UK, and one each in Denmark, France, Germany, Spain, the Netherlands, Finland, China and Australia. Of the top 40 authors, 6 (15%) are female. The top 10 individuals received their Ph.D. from illustrious institutions, 5 in the USA and 5 in the UK: Harvard University, University College London, University of Cambridge (3), University of Illinois, University of Southern California, Case Western University, City University, and Rutgers University. The full list of degree granting institutions includes Columbia University, MIT, Princeton University and Stanford University. Five of the top 40 received their Ph.D. from the University of California at Berkeley.

Many of these authors' work has had a significant impact on the IR field (e.g., Salton G [the SMART system], Roberston SE [probabilistic retrieval model], Van Rijsbergen CJ [IR models], Belkin

NJ [IR evaluation], and Jones SK [TF/IDF—inverse document frequency]) or related fields (Abiteboul S [database management systems], Smith JR [multimedia retrieval, MPEG], Codd EF [OLAP relational model], Ullman JD [database management systems], Zadeh LA [fuzzy logic], Borgman CL [scholarly communication], and Kohonen T [neural networks]). Many also served as program committee members for the SIGIR conferences at some point during the period 1997-2008: Robertson SE, Van Rjisbergen CJ, Spink A, Harman D (chair of TREC), and Voorhees EM (chair of TREC). Some of those coming from related fields served as program committee members for SIGIR (e.g., Smith JR and Ellis D) or related conferences (e.g., SIGMOD [Stonebraker M], and VLDB [Abiteboul S]).

Table 6. Top 40 most prestigious and most popular authors, 1956-2008

| Author (F=Female) | Popularity | Prestige | PC | Ph.D. | Affiliation | Awards* | Key Contribution |
|---|---|---|---|---|---|---|---|
| SALTON G (1927-1995) | 1-1-1-1 | 1-1-1-1 | | 1958 (Harvard) | Cornell Univ. | ASIS&T Award of Merit (1989), Gerard Salton Award (1983)**, ASIS&T Best Book Award (1975) | SMART system |
| ROBERTSON SE | 12-2-6-3 | 10-2-3-4 | 6 | 1976 (Univ. College London, UK) | Microsoft research Cambridge Univ., UK | Tony Kent Strix Award (1998), Gerard Salton Award (2000) | Probabilistic searching model |
| VANRIJSBERGEN CJ | 8-4-4-16 | 10-4-5-6 | 6 | 1972 (Cambridge, UK) | Univ. of Glasgow, UK | Tony Kent Strix Award (2004), Gerard Salton Award (2006) | IR theoretical models |
| BELKIN NJ | 206-10-3-8 | NA-12-2-2 | 5 | 1977 (City, London, UK) | Rutgers Univ. | ASIS&T Award of Merit (2003), ASIS&T Research Award (1997) | IR evaluation |
| RUI Y | NA-NA-194-2 | NA-NA-196-13 | 0 | 1998 (Illinois) | Microsoft China | | Image processing |
| SARACEVIC T | 64-51-5-13 | 71-64-7-3 | 3 | 1970 (Case Western Reserve) | Rutgers Univ. | Gerard Salton Award (1997), ASIS&T Award of Merit (1995) | Digital library |
| CROFT WB | 136-6-7-39 | NA-6-4-7 | 5 | 1979 (Cambridge, UK) | Univ. Massachusetts, Amherst | Gerard Salton Award (2003), ASIS&T Research Award (2000), | Query processing |
| ABITEBOUL S | NA-52-2-7 | NA-NA-17-35 | 0 | 1982 (Southern California) | INRIA, France | SIGMOD Innovation Award (1998) | Database |
| SPINK A (F) | NA-NA-22-6 | NA-NA-26-5 | 8 | 1993 (Rutgers) | Queensland Univ. of Technology, Australia | | Information Seeking behavior |
| JONES KS (F) (1935-2007) | 7-7-15-21 | 2-3-6-10 | 3 | 1964 (Cambridge, UK) | Cambridge Univ., UK | Gerard Salton Award (1998), ASIS&T Award of Merit (2002) | Inverse document frequency (IDF) |
| SMITH JR | NA-NA-49-5 | NA-NA-144-8 | 2 | 1997 (Columbia) | IBM T.J. Watson Research Center | | multimedia retrieval (MPEG-7, MPEG-21) |
| FALOUTSOS C | NA-244-13-7615 | NA-NA-48-18 | 0 | 1987 (Toronto, Canada) | Carnegie Mellon Univ. | | Multimedia retrieval |
| HARMAN D (F) | NA-421-10-19 | NA-NA-18-9 | 7 | N/A | NIST | Tony Kent Strix Award (1999) | Managing TREC |
| VOORHEES EM (F) | NA-253-65-9 | NA-790-103-23 | 7 | 1986 (Cornell ) | NIST | | TREC |
| FLICKNER M | NA-NA-37-11 | NA-NA-383-21 | 0 | N/A | IBM Almaden Research Center | | Image retrieval (QBIC-Query by Image Content system) |
| CODD EF (1923-2003) | 31-8-17-128 | 31-13-21-94 | 0 | 1965 (Michigan) | IBM | Turing Award (1981) | OLAP Relational model |
| FUHR N | NA-584-32-17 | NA-NA-30-17 | 6 | 1986 (Technical Univ. Darmstadt, Germany) | Univ. Duisburg-Essen, Germany | | Digital library, retrieval models |
| JAIN AK | NA-5595-64-12 | NA-NA-829-30 | 0 | 1973 (Ohio State) | Michigan State Univ. | IEEE Technical Achievement Award (2003) | Pattern recognition |
| BATES MJ (F) | 901-39-9-38 | 424-38-8-11 | 2 | 1972 (UC Berkeley) | UCLA | ASIS&T Award of Merit (2005), ASIST Research Award (1998) | Info seeking behavior |
| CHANG SK | 1005-62-11-65 | NA-130-32-31 | 0 | 1969 (UC Berkeley) | Univ. Pittsburg | | Image retrieval |
| ULLMAN JD | NA-20-8-191 | NA-216-34-137 | 0 | 1966 (Princeton) | Stanford Univ. | | Database |
| ZADEH LA | 40-12-58-30 | 92-9-35-62 | 0 | 1949 (Columbia) | UC Berkeley | IEEE Medal of Honor, ACM Fellow | Fuzzy logic |
| BORGMAN CL (F) | NA-18-12-88 | NA-17-11-14 | 4 | 1984 (Stanford) | UCLA | ASIS&T Best Book Award (2008, 2001), | Bibliometrics |
| SWAIN MJ | NA-NA-72-14 | NA-NA-266-28 | 0 | 1990 (Rochester) | | | Image retrieval |
| COOPER WS | 6-11-20-121 | 4-10-10-16 | 0 | 1964 (UC Berkeley) | UC Berkeley | Gerard Salton Award (1994) | IR evaluation |
| STONEBRAKER M | 1480-15-14-167 | NA-50-25-58 | 0 | 1971 (Michigan) | UC Berkeley | John von Neumann Medal, SIGMOD E. F. Codd award | Database management systems |
| PENTLAND A | NA-NA-45-22 | NA-NA-114-14 | 0 | 1982 (MIT) | MIT | | Human-computer interaction |
| INGWERSEN P | NA-104-19-14193 | NA-131-24-19 | 7 | 1991 (Copenhagen Business Univ., Denmark) | Royal School of LIS, Denmark | Derek de Solla Price Medal, ASIS&T Research Award (2003) | Information seeking |
| BAEZA-YATES R | NA-NA-593-4 | NA-NA-2992-91 | 6 | 1989 (Waterloo, Canada) | Yahoo! Research, Spain | | Book: Modern Information Retrieval (1999) |
| BOOKSTEIN A | 25-3-30-277 | 5-5-9-40 | 0 | 1969 (New York) | Univ. Chicago | ASIS&T Research Award (1991) | Indexing theory |
| SMEULDERS AWM | NA-NA-3254-10 | NA-NA-NA-147 | 0 | 1983 (Leiden, Holland) | Univ. Amsterdam, Holland | | Medical retrieval |
| LANCASTER FW | 2-13-28-256 | 3-11-12-23 | 0 | N/A | Univ. Illinois | ASIS&T Award of Merit (1988), ASIST Best Book Award (1992) | Online IR and evaluation |
| SWANSON DR | 27-25-26-94 | 19-15-13-12 | 0 | 1952 (UC Berkeley) | Univ. Chicago | ASIS&T Award of Merit (2000) | Medical IR |
| KOHONEN T | BA-316-87-23 | NA-NA-252-65 | 0 | 1962 (Helsinki Univ. of Technology, Finland) | Helsinki Univ. Technology, Finland | Numerous prizes and awards from IEEE and other organizations for work in AI and neural networks | Neural networks |
| CHEN HC | NA-NA-37-36 | NA-NA-35-32 | 0 | 1989 (New York) | Univ. Arizona | Various awards for MIS-elated work | Data and knowledge mining |
| MARCHIONINI G | NA-261-25-62 | NA-NA-44-33 | 6 | 1981 (Wayne State) | Univ. North Carolina | ASIS&T Research Award (1996) | Human-computer interaction |
| ELLIS D | NA-107-22-71 | NA-689-20-20 | 2 | 1996 (MIT) | Columbia Univ. | | Signal processing |
| FAGIN R | NA-146-42-53 | NA-303-147-89 | 0 | 1973 (UC Berkeley) | IBM Almaden Research Center | IBM Outstanding Technical Award | Schema mapping |
| MANJUNATH BS | NA-NA-269-20 | NA-NA-283-42 | 0 | 1991 (Southern California) | Univ. California, Santa Barbara | | Image processing |

Note: PC indicates membership of the SIGIR program committee for one or more of the years 1997-2008 with the exception of 2003, which we could not find on the website.
*We do not pretend that this list of awards is comprehensive.
** ACM/SIGIR Award for Outstanding Contributions to Information Retrieval as it was known originally.

Table 7 displays several of the major awards in information retrieval and the broader information science field: the Gerard Salton Award, the Tony Kent Strix Award, the ASIS&T Award of Merit, the ASIS&T Research Award and the ASIS&T Best Book Award. For the period 2001-2008, researchers ranked high in prestige have a stronger presence among the award winners than those ranked high on popularity. In the case of first authors of the ASIS&T Best Book Award, only one appears in the list of the most popular authors for the period 2001-2008, while seven are featured on the list of the most prestigious authors. All the Gerard Salton Award winners, with the exception of Cleverdon CW and Dumais S, are included in Table 6. If Table 6 had listed the most prestigious rather than the most popular authors, Cleverdon would have been included because his prestige rank (**6-20-23-**39) is higher than his popularity rank (**3-37-**81-256). The 2009 Gerard Salton Award winner was Susan Dumais from Microsoft Research. She is ranked 80th on prestige and 121st on popularity for the years 2001-2008. Her relatively low ranking may have to do with the fact that she works in industry, with the result that her work may not appear so often in the open literature. She has a higher prestige than popularity ranking, which suggests that domain experts are cognizant of her work. The Gerard Salton Award has nine winners to date, six of whom (67%) were among the top 10 most prestigious authors and only two (22%) among the top 10 most popular authors for the period 2001-2008. This seems to suggest that an author's prestige ranking is a better reflection of perceived scholarly significance than his popularity ranking.

Table 7. Awards for most popular and most prestigious authors

| | Gerard Salton Award | Tony Kent Strix Award | ASIS&T Award of Merit | ASIS&T Research Award | ASIS&T Best Book Award |
|---|---|---|---|---|---|
| Total Awardees | 9 | 9 | 44 | 20 | 36 |
| Top 40 Most Popular Authors, 2001-2008 | 6 | 3 | 5 | 4 | 1 |
| Top 40 Most Prestigious Authors, 2001-2008 | 8 | 3 | 7 | 7 | 7 |

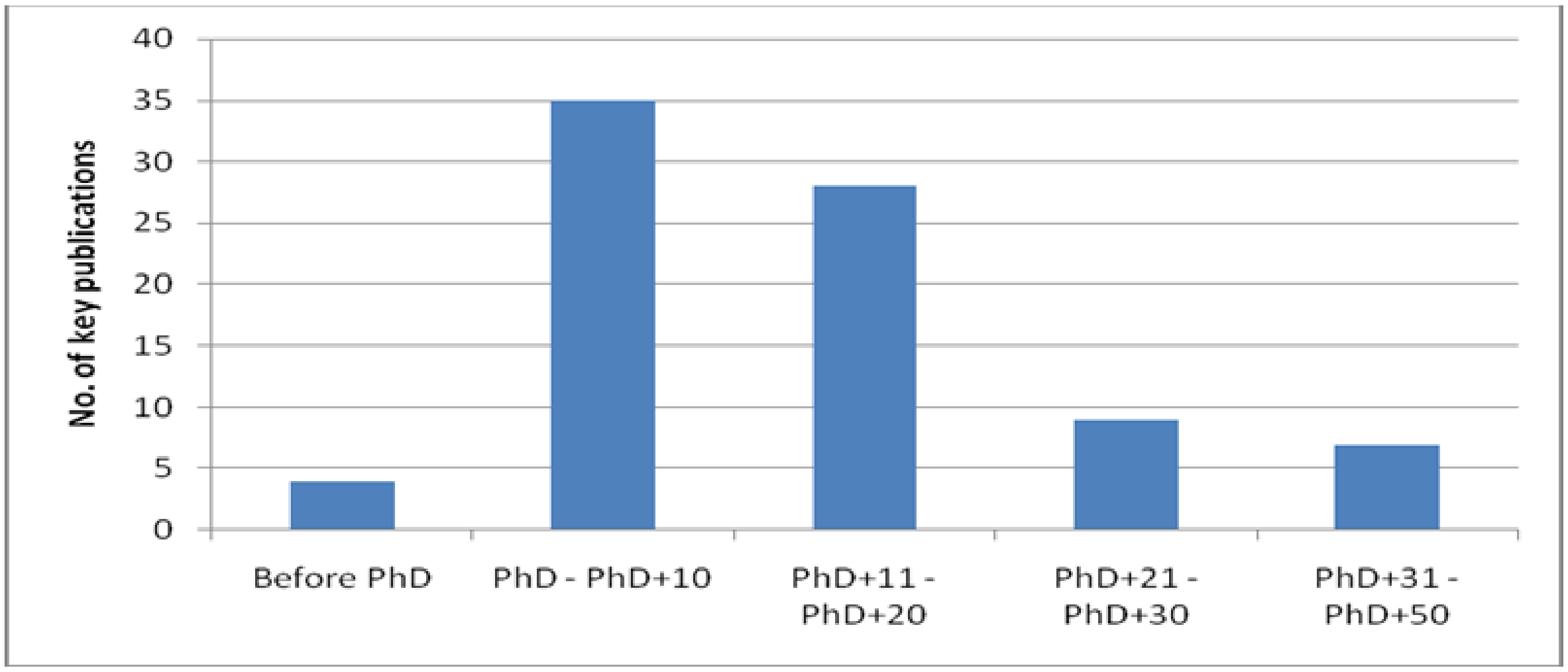

Figure 5. Time before/after award of Ph.D. and production of key publications

We gathered data on when authors produced their most important works (see Figure 5). As mentioned earlier, we defined key publications as those that had been cited at least 40 times. We also determined the date when authors were awarded their doctorate (three did not have a terminal degree). Figure 5 shows that the majority of key publications were produced 10-20 years post-Ph.D., a finding that is congruent with Cronin and Meho's (2007) results. Three of these were books and all three appeared in the popularity column. A comparison of the 10 most highly cited publications for the period 2001-2008 based on popularity and prestige found that only three articles were the same (see Table 8). This further suggests that measures of popularity and prestige are not interchangeable.

Table 8. Ten most highly cited publications, 2001-2008, based on popularity and prestige

| **Top 10 publications based on popularity** | | | | | **Top 10 publications based on prestige** | | | | |
|---|---|---|---|---|---|---|---|---|---|
| Author | Year | Source | Vol/ Book | Citation | Author | Year | Source Field | Vol / Book | Citation |
| SMEULDERS AWM | 2000 | IEEE T PATTERN ANAL | V22 | 368 | SARACEVIC T | 1988 | J AM SOC INFORM SCI | V39 | 78 |
| SWAIN MJ | 1991 | INT J COMPUT VISION | V7 | 312 | BELKIN NJ | 1982 | J DOC | V38 | 65 |
| SALTON G | 1983 | INTRO MODERN INFORMA | Book | 279 | ROBERTSON SE | 1976 | J AM SOC INFORM SCI | V27 | 65 |
| BAEZA-YATES R | 1999 | MODERN INFORMATION R | Book | 263 | SWAIN MJ | 1991 | INT J COMPUT VISION | V7 | 61 |
| DEERWESTER S | 1990 | J AM SOC INFORM SCI | V41 | 230 | SARACEVIC T | 1975 | J AM SOC INFORM SCI | V26 | 59 |
| SALTON G | 1989 | AUTOMATIC TEXT PROCE | Book | 214 | SALTON G | 1990 | J AM SOC INFORM SCI | V41 | 54 |
| RUI Y | 1998 | IEEE T CIRC SYST VID | V8 | 214 | INGWERSEN P | 1996 | J DOC | V52 | 53 |
| PORTER MF | 1980 | PROGRAM | V14 | 208 | SALTON G | 1988 | INFORMATION PROC | V24 | 52 |
| SALTON G | 1988 | INFORMATION PROCESSI | V24 | 200 | DEERWESTER S | 1990 | J AM SOC INFORM SCI | V41 | 46 |

## 6. Summary and concluding remarks

Citation analysis is an established means of assessing the relative impact of a scholar's research. We have described here a novel approach to citation-based evaluation of individuals that factors into account the quality of the papers that cite an author's oeuvre. We measured the prestige of a scholar's work in terms of citations coming from relatively highly cited papers and popularity in terms of citations from all other papers. We used information retrieval as our test site and gathered all IR papers for the years 1956 to 2008 to create our corpus. We broke the analysis down into four time bands and calculated the top 40 authors based on popularity and prestige for each period. We also gathered biographical (e.g., gender) and professional (e.g., organizational affiliation) data on our sample.

The popularity rankings changed over time. Only four scholars managed to maintain a presence in the top 40 rankings for the entire period. The churn rate from one phase to the next was very roughly 40%. Most authors ranked within the top 40 for a single phase; a few for two or three. Rankings based on prestige were more stable than those for popularity. Ten authors ranked in the top 40 for prestige across all four phases. Authors who ranked high on prestige tended to keep their status for 20 or 30 years. We found that authors can rank high on prestige but not on popularity, and vice versa.

Many of the 40 highly ranked authors were affiliated with prestigious organizations—universities and corporate labs in the main—and had received their Ph.D. degrees from leading universities. They were likely to have received awards and honors from the professional community. Six of the nine Gerard Salton Award winners belonged to the top 10 most prestigious authors, and only two were among the top 10 most popular authors for the years 2001-2008. Six females featured among the top 40 ranking authors. Typically, the top-ranked IR scholars produced their key publications approximately 10 to 20 years after completing their doctorate.

Simple citation counting has been a standard approach in first generation bibliometric research. But authors' behaviors (e.g., citing each other or publishing together) generate various kinds of scholarly networks, for example, a paper-citation network, co-authorship network, or author co-citation network. The topology of these social network graphs should not be ignored in assessing the impact of a scholar's research. For example, in a co-authorship network, authors with direct or indirect links to author A will transfer their weight to this author. Simple citation counting only calculates the number of nodes with direct links without considering the weights transferred by indirect nodes.

Both HITS (viewed as a precursor of PageRank) and PageRank use link analysis algorithms that take the link graph topology into consideration when rating web pages. When ranking one node in a graph, they consider the weights coming from not only directly linked nodes but also indirectly linked nodes. The basic premise is that "the creator of page p, by including a link to page q, has in some measure conferred authority on q" (Kleinberg, 1998, p. 2). HITS takes into account both hub and authority; for example, the web page www.harvard.edu should have the highest authority for Harvard University. Hubs are those web pages linking to related authorities, such as web pages with large directories, that led users to other authorized pages, for example, www.dmoz.org (the Open Directory Project). PageRank is very similar to HITS and uses random surfer theory to predict the possibility of any given web page being visited. The PageRank formula consists of two parts: simple counting of nodes (similar to simple citation counting) and weight transfer based on graph topology. A damping factor is used in the formula to balance these two parts. By tuning the damping factor, emphasis can be placed on either of the two parts. For example, if the damping factor is set at low, simple node counting will play a major role in determining the PageRank score, and vice versa (Ding et al., 2009).

The weighted citation counting approach being proposed here demonstrates the value of adding weights to citations so that papers cited by highly cited papers receive more weight than those cited by non-highly cited papers. However, it does not consider the graph topology of citation networks. Several researchers have shown that PageRank can capture the prestige of journals (Bollen, Van de Sompel, Hagberg, & Chute, 2009; Leydesdorff, 2009; Franceschet, 2009), but very few, if any, have tested this at either the author or paper level. We plan to apply the model described here to the paper level and further test the PageRank and HITS algorithms to identify novel methods for measuring popularity and prestige.

## Acknowledgment

The authors would like to thank Johan Bollen and two anonymous referees for their insightful comments on an earlier draft.